# Topology and the Web of Twentieth Century Science

J. C. Phillips

Dept. of Physics and Astronomy, Rutgers University, Piscataway, N. J., 08854-8019

**Scientific communication is an essential part of modern science: whereas Archimedes worked alone, Newton (1676) acknowledged that "If I have seen a little further, it is by standing on the shoulders of Giants." How is scientific communication reflected in the patterns of citations in scientific papers? How have these patterns changed in the 20th century, as both means of communication and individual transportation changed rapidly, compared to the earlier post-Newton 18th and 19th centuries? Here we discuss a physical model for scientific communications, based on an informetric study of 25 million papers and 600 million citations; the physical model itself relies on analogies with glass relaxation, where virtually identical patterns have been identified in 50 well designed experiments. The model reveals a surprisingly universal internal structure in the development of scientific research, which is essentially constant across the natural sciences, but which changed qualitatively around 1960.**

Early informetric studies of citation patterns identified power law patterns, which were interpreted in terms of a cumulative advantage model, but more recent studies have shown a mixed distribution, with power laws prevailing at larger numbers of citations and stretched exponentials giving better fits for smaller numbers [1]. A recent very large scale study of 20th century citations [2] found that stretched exponentials give better fits to citation chains for both low and intermediate citation levels n < $n_1$, where $n_1$ ~ 40 earlier in the century, and $n_1$ ~ 200 later. The parameters involved in the stretched exponentials appear to follow a pattern already recognized in glassy relaxation, an unexpected result which suggests a physical model for scientific communications.

Stretched exponential relaxation (SER) was first recognized by Kohlrausch in 1854 as providing a very good fit (for example, far superior to two exponentials, although the latter involve four parameters) to the residual decay of charge on a Leyden jar. The stretched exponential function $A\exp(-(t/\tau)^\beta)$ contains only three parameters, with $\tau$ being a material-sensitive parameter and $0 < \beta < 1$ behaving as a kinetic exponent. SER is regarded by some as the oldest unsolved problem in science, and it is closely related to the (possibly insoluble) mathematical problem of proving the existence of a closest particle packing density for general interparticle forces.

It was long believed that the ability of stretched exponentials to give better fits to data was accidental, but as sample homogeneity and data improved, so did the accuracy of SE three parameter fits! Finally in the 1970's rigorous derivations of SER emerged [3]. It turns out that the key to understanding SER, which is characteristically observed most accurately in microscopically homogeneous glasses, is to recognize that such glasses, unlike most crystals and normal liquids, are not simply connected; instead they are multiply connected. During the quenching process, a good glass-former avoids crystallization, and it may appear to be microscopically homogeneous, but it actually contains quenched-in, randomly distributed defects that function as traps for diffusive electronic or molecular excitations. When excitations decay, they diffuse to these traps and disappear. The excitations that are closest to the traps disappear first, and the excitations that are further away take longer times to reach the traps. It is this effect (a kind of memory effect, characteristic of non-equilibrium systems) that produces SER.

The success of the SLP model for $\beta$ [3] is restricted to glasses. Fortunately there are many glasses, especially network glasses (like silica and window glass) and molecular glasses (like glycerol) where accurate temporal relaxation data are abundant (at least 50 well-studied cases) and the microscopic theory based on excitations diffusing to traps has proved to be extremely accurate, explaining multiple concordances in the modern data. In fact, the accuracy is so great that the odds against the theory (which cannot be rigorously proved) being successful by accident are of order $10^{50}$ to 1 [4,5]. This success for such a wide variety of materials can be traced to a common characteristic of glasses, which is that they fill space nearly optimally, with the crystalline phase being regarded as a singularity.

The key result of the diffusion to traps model is that in the glass $\beta_g = d/(d + 2)$, where d is the effective dimensionality of configuration space. For short range forces d = 3 and $\beta_g = 3/5$, a result confirmed by both numerical simulations and many experiments [3,4]. However, the long-chain nature of polymers introduces long-range elastic forces, and in some other materials (like solid electrolytes) there are strong long-range Coulomb forces. In these long-range cases it was found that $\beta_g = d^*/(d^* + 2)$, with $d^* = fd$, and $f = 1/2$. This result is also very well supported by experiment, but it has not been derived from a mathematical model. Because space-filling itself involves a delicate balance between short-range constraints and long-range attractive forces (for instance, Van der Waals interactions), such an equalized kinetic balance between a short-range interactions (which promote relaxation) and long-range interactions (which are ineffective and do not contribute to relaxation) seems quite plausible. In this case $d^* = 3/2$ gives $\beta_g = 3/7$, a value observed in experiments on microscopically homogenous polymers and solid electrolytes. The crossover from 3/5 to 3/7 has been observed in the case of monomer (PG) to polymer (PPG) propyl glycol [4].

**Informetric Citation Results**

The startling result found in the informetrics study [2] is shown in Fig. 1. Not only do the intermediate distributions exhibit SER, but they also bifurcate in 1960, with stretching exponents $\beta$ that match (within a few %) the $\beta_g = (\beta_{g11}, \beta_{g22}) = (3/5, 3/7)$ values identified IN THE ABSTRACT of [1]: 3/5 holds < 1960, while 3/7 holds > 1960! To be more precise, the fitted values are $\beta_c = (\beta_{c11}, \beta_{c22}) = (0.57, 0.47)$.

**Physical Model**

The physical content of glass relaxation, in terms of either short-range forces only ($\beta_g = \beta_1 = 3/5$), or a balanced mixture ($f = 1/2$) of short- and long-range forces ($\beta_g = \beta_2 = 3/7$), is contained in the multitude of details identified in 50 well designed experiments [3-5]. However, what happens when the sample is disordered but is not a conventional glass?

This question has been elegantly answered in a recent study of luminescence from 16 single-crystal isoelectronic $ZnSe_{1-x}Te_x$ alloys of commercial quality (these alloys are used in orange light-emitting diodes) [6], with results for $\tau(x)$ and $\beta(x)$ shown in Fig. 2. After optical excitation has produced free electrons and holes, there are many channels for radiative recombination. In the pure crystal ($x = 0$ and $x = 1$) the free electron and hole can recombine, or they can first form an exciton (hydrogenic) bound state and recombine, with simple exponential kinetics ($\beta = 1$). In the alloy, $0 < x < 1$, either the electron or the hole or both can be localized, adding further recombination channels. The competition between this multiplicity of unavoidably disordered channels always leads to $\beta < 1$ in the alloys, in other words, SER.

From a first glance at Fig. 2 it might appear that $\beta = \beta(x)$ is a smooth function with no special features. However, closer inspection reveal two special features associated with the extrema of $\tau(x)$ (near $x \sim 0.1$) and $\beta(x)$ ($x \sim 0.25$). With increasing x, Te clusters attracting holes begin to form (the ionization potential of ZnTe is smaller than that of ZnSe). The holes also have much heavier masses than the electrons, which means that the holes can localize at these Te-rich clusters while the electrons remain delocalized, a situation favorable for long lifetimes, which explains the maximum in $\tau(x)$ at $x \sim 0.1$. The localized holes will be strongly disordered, and when the clusters are large enough to localize nearly all the holes, these localized hole states will nearly fill space – in other words, near the maximum in $\tau(x)$ at $x \sim 0.1$, the holes form a glassy state. For $x > 0.1$, the clusters will begin to overlap, and the holes will begin to delocalize, losing the glassy character of the hole distribution.

Very near the maximum in $\tau(x)$, we expect that radiative recombination will be dominated by states in which the electron and hole have formed an exciton state bound to the Te-rich clusters. The radius of this exciton state defines a characteristic length for short-range interactions, and thus the value of $\beta(0.1)$ is close to 3/5, as shown in the Figure. To understand the meaning of the minimum in $\beta(x)$ near $x = 0.25$, which is also a maximum for overall relaxation, we need to know some details of the hole and electron band states of the alloy.

Although the energy gaps at x = 0 and x = 1 are nearly equal, the alloy electron-hole energy gap is strongly bowed downward, and the band edge shifts from 2.7 eV (x = 0) to 2.05 eV for x = 0.35. In 1993 [7] observed a continuous Localized – Extended (L-E) transition from the recombination through free and bound exciton states to the recombination of excitons localized by the compositional fluctuations of the mixed crystal in the concentration region of about x = 0.25. This L-E transition is the two-particle analogue of the better known metal-insulator transition. The transition occurs when the phase spaces accessible to the localized and extended states are nearly equal. This is again a kind of dual space-filling condition, and it is the kind of condition where one expects an equalized glassy balance between short-range interactions (which promote relaxation) and long-range interactions (which are ineffective and do not contribute to relaxation), in other words, f = 1/2, and $\beta(0.25)$ = 3/7. As shown in Fig. 2, this is what actually is observed at the $\beta$ minimum. Similar size-dependent effects have been observed in luminescence from pure compound quantum dots; $\beta = \beta_1$ or $\beta_2$, depending on the size of excitons compared to dot size [4,5].

The discussion of $\tau(x)$ and $\beta(x)$ in these 16 single-crystal $ZnSe_{1-x}Te_x$ alloys may seem schematic, but it is important to realize that it has been carried out, not in isolation, but in the context of 50 other well designed experiments [3-5], which define a remarkably simple framework for describing complex relaxation in multichannel contexts. The parallels between these experiments concerning the distribution of physical relaxation times in glasses and the statistics of low and intermediate citation distribution are striking, both qualitatively and quantitatively.

**Physical Ranges of Citation Interactions**

Is it feasible to compare citation distributions with the distribution of physical relaxation times in glasses? [2] surveyed $6.10^8$ citations, but this number, although very large (much larger than in previous studies of a few million citations [1]), is still small compared to Avogadro's number. Moreover, scientific culture is far from microscopic homogeneity, with significant differences between North and South America, Europe, Africa and Asia. It is very interesting that if we regard $\beta_c$ and $\beta_g$ as the diagonal elements of $\beta_{ij}$ matrices (i,j =1,2) in citation and glass spaces respectively, then $\beta_{c11} + \beta_{c22} = \beta_{g11} + \beta_{g22}$, in other words, the traces of the $\beta_c$ and $\beta_g$ matrices are equal to within 1%.

The citation process can be compared to luminescence of electron-hole pairs, with electrons representing citing papers and holes representing cited papers. In general, radiative electron-hole recombination is inefficient: most pairs do not recombine radiatively, and similarly most papers are not cited in other papers. As we saw in Fig. 2, there are two stable cases of SER, the first arising from a maximum in $\tau$, corresponding to $\beta_{g11} = 3/5$ (short-range forces only), and the second corresponding to $\beta_{g22} = 3/7$ (mixed short-and long-range forces). Thus it is not surprising to find a parallel matrix bifurcation in $\beta_{c11}$ and $\beta_{c22}$.

**Damping Interactions**

Continuing the matrix analogy, we see that to explain the reduction in the splitting of the informetric citation values relative to physical values found in glasses, we must assume that while the glass matrix can be taken as diagonal, the citation matrix is anti-Hermitian, with identical imaginary off-diagonal values $\beta_{c12} = \beta_{c21} = ai$, with $a \ll 1$. Anti-Hermitian matrices arise in the context of a new and very efficient way to calculate two-electron excited states of many-electron atoms and molecules, without doing configuration interaction, by using an anti-Hermitian contracted Schrodinger equation (ACSE) [8]. In view of the well-known parallels between temperature-dependent dynamics for the partition function Z ($dZ/d\gamma = -HZ$, with $\gamma = 1/kT$) and the Schrodinger equation for the time-dependent wave function $\psi$ ($d\psi/dt = iH\psi$), this suggests a way of understanding both citation distributions and glassy relaxation. Given the indubitable concordances between the citation and glassy matrices, such parallels are attractive.

Within the ACSE one calculates two-body charge density matrices that can be said to be the classical analogue of citer-cited pair interactions. The ACSE is an optimized reduction of the enormously complex problem of quantum configuration interaction. In principle the two-body charge density matrices are related to the three- and four- body charge density matrices, which in turn are related to still higher order charge density matrices. Moreover, because interference effects are not included, these classical correlation functions in general do not correspond to actual wave function solutions of the Schrodinger equation itself. Nevertheless, it has been found in many examples that restricting the higher order corrections to the leading anti-Hermitian third-order term yields excellent results, at least an order of magnitude better than obtained from the full (Hermitian + anti-Hermitian) third-order term [8]. Classically speaking, the anti-Hermitian interaction corresponds to damping of the pair interaction by a third particle. Apparently such damping does not occur in glasses, probably because of the space-filling condition. It does occur in citation space (albeit only weakly); it admixes a small fraction of additional long-range interactions into the "pure" short-range case, and vice-versa.

Let us first consider the case of a simple exponential distribution, $\beta = 1$. Physically this corresponds to an isolated excitation in a gas or normal liquid, which decays at a constant rate, independent of history or loss of energy to its environment. This could also be the pattern for papers dominated by self-citations. By contrast, collaborating individuals create a partially self-sustaining environment, in which citations can accumulate to a larger extent for the more successful or appealing papers. There are various "trapping" events that could limit such an accumulation (individual members develop other interests, acquire new equipment, or join other groups), but encounters with these all have a diffusive nature, described mathematically by continuous random walkers.

**Sociology of Natural Sciences**

The physical distinction between short- and long-range forces may have a social analogy. Scientific collaboration occurs on multiple levels, from details to general ideas and motivation. Direct collaboration requires frequent contacts of individuals, and appears quite analogous to short-range interactions (nearest and sometimes second nearest neighbors, but seldom third neighbors) in glasses. The nature of this glassy environment depends on its size. In particular, the apparent dimension of the environment is likely to be d = 3 (latitude, longitude, and time). Since there is a strong tendency to cite papers that one knows best, and these are often the papers of one's colleagues, and possibly their closest colleagues, the citation number for a given paper will be limited largely to this circle. This corresponds to $\beta_1 = 3/5 \sim 0.57$, which gives the best fit to low and intermediate citation levels prior to 1960.

Because the detailed aspects of a research paper are best known only to this small space-time circle, only the more general aspects will give rise to a larger number of citations. After 1960, scientific conferences became popular and grew steadily in size, for a wide variety of practical reasons – an upsurge in funding for science after Sputnik, rapid moderation of commercial jet fares (the first successful commercial jet, the Boeing 707, was delivered in waves, with the first two delivery waves peaking in 1960 and 1967), etc. At scientific conferences one encounters a wider range of ideas in a much less detailed format, which corresponds to long-range interactions, and appears to give a broader distribution of citations. The wider range of ideas can survive longer, the paper can continue to accumulate citations, and because $f \sim 1/2$ the diffusive interactions affect only details, while leaving the larger ideas intact and still capable of accumulating citations. Here it is quite striking that the balance between short- and long-range effects on relaxation in glasses should be so closely echoed (similar values of f!) in the competition between narrower and wider influences on citation patterns.

**Informalogic of Science**

The hardest question has been left for last. Few scientists will be surprised to learn that scientific communications have a diffusive character, or that the "trappy" events that terminate a paper's citations are randomly distributed. While the history of scientific discoveries is often written so as to make them appear inevitable, those closest to the discoveries have often remarked on the surprisingly large part played by accident, a famous case being the discovery of penicillin in bread mold.

The most difficult question then is, why does the distribution of scientific citations follow SER *in glasses* so well, and what does this tell us about science and scientific research? Quite naturally [3-5] emphasized the universality of SER in microscopically homogenous glasses, including polymers. However, current research is showing that polymer blends [9,10] often exhibit multiple SER (not single SER, as shown in Fig.1 here for citations, and in [4] for many microscopically homogeneous glasses, including polymers). Thus there must be internal factors in science that give scientific research a glassy character, and specifically correspond to microscopic homogeneity. A possible candidate for glassy character is the search for novelty, which expands to fill the accessible space densely, just as a glass, subject to stronger direct forces, is compacted by weaker residual forces. A possible candidate for microscopic homogeneity is the insistence that scientific theories be validated by experiments – this is what ultimately limits branching balkanizations such as Lysenkoism.

This leaves one last question. Granted that careful scientific research leads to SER, why does it lead either to $\beta_1$ or to $\beta_2$? The answer to this question may be contained in Fig. 2. There we notice that although there is a wide range of values of $\beta$, the two common values are metastable extrema, either in $\tau$ or in $\beta$. The characteristic feature of extrema is their stability, which in turn favors consensus. While scientists seek originality, they also attempt to place their new results in the context of established results, which is most accomplished with papers that have stabilized their informational contexts. Thus (of course only in retrospect!) the appearance of SER in citation patterns, and even its rather striking change in character around 1960, appears obvious.

## References

1. Redner, S. Citation statistics from more than a century of Physical Review *arXiv Physics:* 0407137 (2004).

2. Wallace, M. L., Lariviere, V. & Gingras, Y. Modeling a century of citation distributions *J. Inform.* **3**, 296 -303 (2009).

3. Phillips, J. C. Stretched exponential relaxation in molecular and electronic glasses *Rep. Prog. Phys.* **59**, 1133-1207 (1996).

4. Phillips, J. C. Slow dynamics in glasses: A comparison between theory and experiment *Phys. Rev. B* **73**, 104206 (2006).

5. Phillips, J. C. Microscopic Aspects of Stretched Exponential Relaxation (SER) *arXiv Physics:* 0903.1067 (2009).

**6.** Lin, Y. C. *et al.,* Time-resolved photoluminescence of isoelectronic traps in ZnSe1-xTex semiconductor alloys *Appl. Phys. Lett.* **93**, 241909 (2008).

7. A. Naumov, H. Stanzl, K. Wolf, S. Lankes & W. Gebhardt, Exciton recombination in Te-Rich ZnSexTe1-x epilayers *J. Appl. Phys.* **74**, 617 (1993).

8. Mazziotti, D. A. Anti-Hermitian Contracted Schrödinger Equation: Direct Determination of the Two-Electron Reduced Density Matrices of Many-Electron Molecules *Phys. Rev. Lett*. **97**, 143002 (2006).

9. Liu WJ, Bedrov D, Kumar SK, Veytsman B. & Colby R. H., Role of Distributions of Intramolecular Concentrations on the Dynamics of Miscible Polymer Blends Probed by Molecular Dynamics Simulation *Phys. Rev. Lett*. **103**, 037801 (2009).

10. Sakai V. G., Maranas J. K., Peral I., & Copley J.R. D., Dynamics of PEO in blends with PMMA: Study of the effects of blend composition via quasi-elastic neutron scattering *Macromol.* **41**, 3701-3710 (2008).

# Figure Captions

Fig. 1. Evolution of the number of papers n with a number of citations $\tau$ (or T) as a function of the decade in which the paper was published. The distributions for the decades up to 1960 are well fitted by an SE with $\beta = 0.57$, while the decades after 1960 are fitted with $\beta = 0.47$. The crossover at 1960 is unambiguous [2], as are the accuracies of the SE fits.

Fig. 2. Annotated data [6] on luminescence in isoelectronic Zn(Se,Te) alloys. The peak in relaxation time $\tau$ occurs near x = 0.10, where there is a break in $d\beta/dx$. These are the maximally localized states, which behave as quasi-particles subject to short-range forces only, with $\beta = \beta_1 = 3/5$. The localized-extended transition occurs at x = 0.22, where the long-and short-range forces are equally weighted, and $\beta = \beta_2 = 3/7$. The $\beta$ fitting errors (0.2%) are about ten times smaller than the size of the data points, as expected from semiconductor samples of commercial homogeneity.

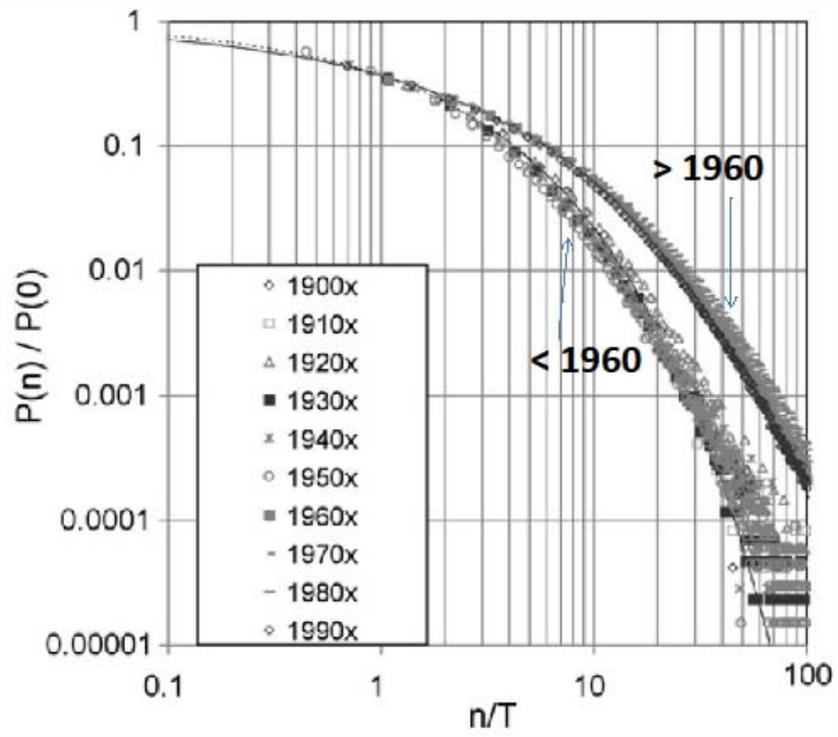

Fig. 1

...

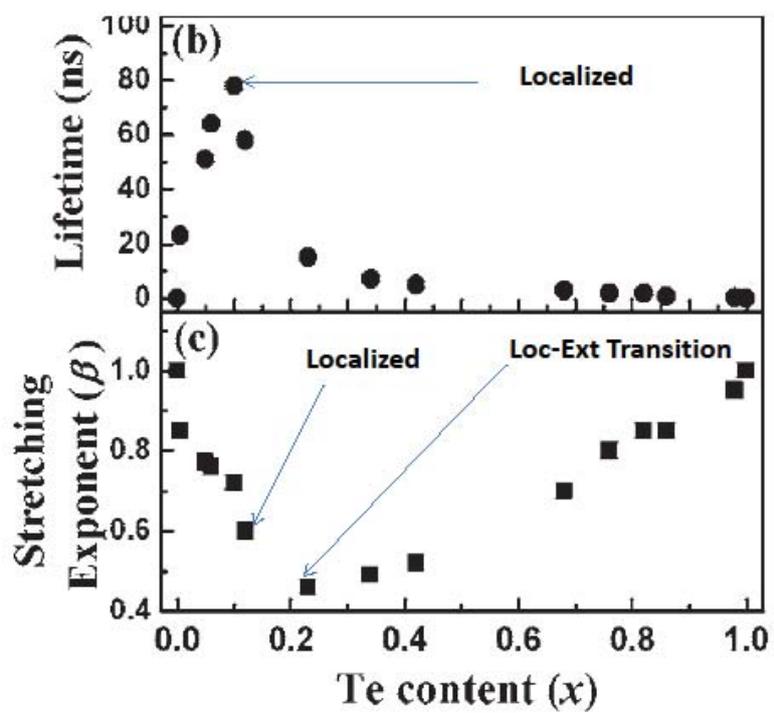

Fig. 2.